\baselineskip=18pt
\def\a{\alpha}\def\b{\beta}\def\c{\chi}\def\e{\epsilon}
\def\f{\phi}\def\g{\gamma}
\def\l{\lambda}\def\m{\mu}\def\n{\nu}\def
\p{\pi}\def\r{\rho}\def\s{\sigma}
\def\y{\eta}\def\x{\xi}\def\z{\zeta}

\def\D{\Delta}\def\G{\Gamma}\def\L{\Lambda}
\def\O{\Omega}\def\S{\Sigma}

\def\na{\nabla}
\def\inf{\infty}\def\id{\equiv}\def\mo{{-1}}\def\ha{{1\over 2}}

\def\const{{\rm const}}

\def\mn{{\mu\nu}}

\def\fe{field equations }\def\bh{black hole }\def\as{asymptotically }
\def\coo{coordinates }

\def\cosc{cosmological constant }

\def\ssy{spherically symmetric }
\def\cp{critical points }

\def\sch{Schwarzschild }\def\ads{anti-de Sitter }

\def\KK{Kaluza-Klein }\def\des{de Sitter }

\def\GR{general relativity }
\def\GB{Gauss-Bonnet }

\def\ab{asymptotic behavior }

\def\section#1{\bigskip\noindent{\bf#1}\smallskip}

\def\PL#1{Phys.\ Lett.\ {\bf#1}}
\def\PRL#1{Phys.\ Rev.\ Lett.\ {\bf#1}}
\def\PR#1{Phys.\ Rev.\ {\bf#1}}\def\CQG#1{Class.\ Quantum Grav.\ {\bf#1}}
\def\NP#1{Nucl.\ Phys.\ {\bf#1}}
\def\JMP#1{J.\ Math.\ Phys.\ {\bf#1}}

\def\MPL#1{Mod.\ Phys.\ Lett.\ {\bf #1}} 
\def\PRep#1{Phys.\ Rep.\ {\bf#1}}

\def\grq#1{{\tt gr-qc/#1}}

\def\ref#1{\medskip\everypar={\hangindent 2\parindent}#1}
\def\beginref{\begingroup
\bigskip
\centerline{\bf References}
\nobreak\noindent}
\def\endref{\par\endgroup}

\def\er{{\cal R}}\def\es{{\cal S}}\def\ef{e^{2\f}}
\def\eg{{\cal G}}
\def\qua{{(4)}}\def\np{{(n+4)}}

{\nopagenumbers \line{\hfil 15 September 2006}
\vskip80pt
\centerline{\bf Black hole solutions of dimensionally reduced
Einstein-Gauss-Bonnet gravity}
\centerline{\bf with a cosmological constant}

\vskip40pt
\centerline{{\bf M. Melis}\footnote{$^\dagger$}{e-mail:
maurizio.melis@ca.infn.it} and
{\bf S. Mignemi}\footnote{$^\ddagger$}{e-mail:
smignemi@unica.it}}
\vskip10pt
\centerline {Dipartimento di Matematica, Universit\`a di Cagliari}
\centerline{viale Merello 92, 09123 Cagliari, Italy}
\centerline{and INFN, Sezione di Cagliari}
\vskip100pt
\centerline{\bf Abstract}
\vskip10pt
{\noindent
We study the phase space of the \ssy solutions of the system obtained
from the dimensional reduction of the six-dimensional Einstein-\GB action
with a cosmological constant.
We show that all the physical solutions have \ads asymptotic behavior.}
\vskip100pt\
P.A.C.S. Numbers: 97.60.Lf 11.25.Mj
\vfil\eject}

\section{1. Introduction.}

In a recent paper [1], one of us has investigated the black hole solutions of a
dimensionally reduced gravity model with \GB (GB) corrections to the Einstein-Hilbert (EH)
lagrangian of general relativity. Using global methods it was possible to
classify the solutions in terms of their asymptotic behavior, in analogy with
what done in [2] in the pure Einstein-Hilbert case and in [3] in the presence of a
cosmological constant (CC).
The results of the work of [1] showed a remarkable similarity between
the EH-GB and the EH-CC system. It is therefore interesting
to extend that investigation to the case where both the GB term and a cosmological constant
are added to the EH lagrangian.

As it is well known, GB terms arise as a natural extension to higher dimensions of the
EH lagrangian, sharing most properties of \GR [4]. They have therefore found many
applications in Kaluza-Klein theories [5].
Of course, the properties of \bh solutions in  dimensionally reduced models are of great
interest. In the case of pure Einstein gravity they were studied in [2], where
it was shown that the only solution of physical interest is the
four-dimensional \sch metric with flat internal space.
When a \cosc is added, physically reasonable solutions have \ads
asymptotics and negative-curvature internal space [3]. The case of a six-dimensional
EH-GB model without \cosc was studied in [1]. It displayed some similarities with the case
of EH-CC, in particular the existence of \as\ads black holes\footnote{*}{Black holes in
higher-dimensional EH-GB theories admitting higher-dimensional spherical symmetry have been
extensively studied in the literature [6].}.

In this paper, our aim is to extend these investigations to the case of the EH-CC-GB
system, classifying all the solutions having the form of a direct product of a four-dimensional
\ssy \bh with a maximally symmetric internal space.
Since a general discussion would be too involved, we shall limit ourselves
to the case of six dimensions, where the only relevant GB correction is quadratic in the
curvature and has the form $\es=\er^{\m\n\r\s}\er_{\m\n\r\s}-4\er^{\m\n}\er_{\m\n}+\er^2$.

In ref.\ [1-3], this topic was studied by considering
the phase space of the solutions of the field equations.
It is well known that when a GB term is added to the action, the field equations remain second
order and linear in the second derivatives, but no longer quadratic in the first derivatives.
This fact gives rise to several technical problems. In particular, the potential of the dynamical
system is no longer polynomial, but presents poles for some values of the variables.

The result of our investigation is that all the relevant solutions have \ads asymptotic behavior,
as in the EH-CC case. However, in our case the possibility emerges of a flat or
positive-curvature internal space.


Let us consider the $\np$--dimensional action
$$I=\int\sqrt{-g}\ d^\np x\ (2\b+\er^\np+\a\es^\np),\eqno(1.1)$$
where $\er^\np$ is the curvature scalar, $\es^\np$ the quadratic GB
term of the manifold, $2\b$ the cosmological constant, and $\a$ a coupling
parameter of dimension $[L]^2$.

We perform a dimensional reduction which casts the metric in the form of a direct product
of a four-dimensional manifold with an $n$-dimensional space of constant curvature, whose
size is parametrized by a scalar field $\f$.
As discussed in [1],  it is not possible to find an ansatz for
the metric of the EH-GB system that completely disentangles the scalar field $\f$
from the curvature in the dimensionally reduced action, except when the internal space
is flat. Therefore we maintain the usual ansatz
$$ds_\np^2=e^{-n\f}ds_\qua^2+e^{2\f}g_{ab}^{(n)}dx^adx^b,\eqno(1.2)$$
where $ds_\qua^2$ is the line element of the four-dimensional spacetime  and $g_{ab}^{(n)}$
is the metric of the $n$-dimensional maximally symmetric internal space, with
$\er_{ab}^{(n)}=\l_ig_{ab}^{(n)}$.
The action is dimensionally reduced to
$$\eqalignno{&I=\int\sqrt{-g}\ d^4x\Big[(1+2\a\l_i\,e^{-2\f})\er^\qua+\a\,e^{n\f}\es^\qua+
4n\a\,e^{n\f}\eg_\mn^\qua\na^\m\phi\na^\n\phi&\cr
&+\left({n(n+2)\over2}-(n^2-2n-12)\a\l_i\,e^{-2\f}\right)(\na\f)^2-{n(n+2)(n^2+n-3)\over3}
\ \a\, e^{n\f}(\na\f)^4&\cr
&+2\b\,e^{-n\f}+\l_i\,e^{-(n+2)\f}+(n-2)(n-3)\a\l_i^2e^{-\np\f}\Big].&(1.3)}$$

The ground state is assumed to have the form of a direct product of a four dimensional
and an $n$-dimensional maximally symmetric space, i.e.\ $\er^\qua_{\m\n\r\s}=\L_e(g^\qua_{\m\r}g^\qua_{\n\s}-g^\qua_{\m\s}g^\qua_{\n\r})$,
$\er^{(n)}_{\m\n\r\s}=\L_i(g^{(n)}_{\m\r}g^{(n)}_{\n\s}-g^{(n)}_{\m\s}g^{(n)}_{\n\r})$.
Substituting this ansatz into the \fe derived from (1.1), one obtains
$$\eqalignno{&\a[(n-1)(n-2)(n-3)(n-4)\L_i^2+24\L_e^2+24(n-1)(n-2)\L_e\L_i]+(n-1)(n-2)\L_i+12\L_e+2\b=0,&\cr
&\a[n(n-1)(n-2)(n-3)\L_i^2+12n(n-1)\L_i\L_e]+n(n-1)\L_i+6\L_e+2\b=0.&(1.4)}$$
In the case of interest, $n=2$, if $\a\b\le3/4$ and $\a\b\ne5/12$, the system admits two solutions
$$\L_e={1\over4\a}\left(-1\pm\sqrt{1-{4\a\b\over3}}\right),\qquad
\L_i={3\left(-1\pm\sqrt{1-{4\a\b\over3}}\right)-4\a\b\left(-1\pm3\sqrt{1-{4\a\b\over3}}\right)
\over4\a(5-12\a\b)}.\eqno(1.5)$$
If $\a>0$ and $\b>0$, the values of $\L_e$ are both negative, corresponding to \ads spacetime,
if $\a<0$ and $\b<0$, both positive (\des spacetime); finally, if $\a$ and $\b$ have opposite sign
one solution is positive and one negative. The values of $\L_i$ are both positive, corresponding to
internal space $S^2$, if $\a>0$, and $\a\b>5/12$ or $\a<0$ and $\a\b<5/12$, both negative
(internal space $H^2$) otherwise.

Consequently, for a range of values of $\a$ and $\b$ \bh solutions of (1.1) may have \ads behavior at
spatial infinity (we are not interested in de Sitter spacetimes since they do not have an asymptotic
region).
In the limit $\b\to0$ one recovers the solutions with vanishing cosmological constant of ref.\ [1].
The limit $\a\to0$ is instead singular: in absence of GB corrections the unique solution of (1.4) is
given by $\L_e=-\b/6$,\ $\L_i=-\b/2$ and is therefore $AdS\times H^2$ for $\b>0$, or $dS\times S^2$
for $\b<0$.
Another interesting limit case arises when $\a\b=3/4$. In this limit the internal space is flat.

\bigbreak
\section{2. The dynamical system.}

In [1] the dynamical system associated to the \ssy solutions of the model was derived
when $\b=0$. We now review that derivation when a cosmological constant is added to the
lagrangian.
For the four-dimensional metric we adopt the \ssy ansatz [1-3]
$$ds_\qua^2=-e^{2\n}dt^2+\s^{-2}e^{4\z-2\n}d\x^2+e^{2\z-2\n}g_{ij}dx^idx^j,\eqno(2.1)$$
where $\n$, $\z$ and $\s$ as well as $\f$ are functions of $\x$ and $g_{ab}$ is the metric
of a two-dimensional maximally symmetric space, with $\er_{ij}=\l_eg_{ij}$. Of course,
in the case of physical interest, $\l_e>0$.

Defining the new variables [1]
$$\c=2\z-\n-\f,\qquad\y=2\z-\n-2\f,\eqno(2.2)$$
and substituting the ansatz (1.2), (2.1) into the action, after factoring out the
internal space the action can be cast in the form
$$\eqalignno{
I=\, -8\p&\int d^4x\bigg\{\s\,\Big[6\c'^2+3\z'^2+3\y'^2-8\c'\z'-8\c'\y'+4\z'\y'\Big]-
{1\over\s}(\l_ee^{2\z}+\l_ie^{2\y}+\b e^{2\c})&\cr
&+4\a e^{-2\c}\Big[\s(\y'-\c')(4\z'+3\y'-5\c')\l_ee^{2\z}+
\s(\z'-\c')(3\z'+4\y'-5\c')\l_ie^{2\y}&\cr
&-\s^3(\z'-\c')(\y'-\c')(11\c'^2+4\z'^2+4\y'^2+7\z'\y'-13\c'\z'-13\c'\y')-\l_e\l_i\,
{e^{2(\z+\y)}\over\s}\, \Big]\bigg\}.&(2.3)}
$$

As usual, the action (2.3) does not contain derivatives of $\s$, which
acts therefore as a Lagrangian multiplier enforcing the Hamiltonian
constraint. Moreover, in spite of the presence of the higher-derivative
GB term,
it contains only first derivatives of the fields, although up to the fourth power, and
therefore gives rise to second order field equations. Finally,
the action is invariant under the interchange of $\z$ and $\y$.

One can now vary (2.3) and then write the \fe in first order form in terms of
the new variables,
$$W=\c',\quad X=\z',\quad Y=\y',\quad U= e^\c,\quad Z=\ e^{\z},\quad V=\ e^{\y},\eqno(2.4)$$
which satisfy
$$U'=WU,\qquad Z'=XZ,\qquad V'=YV.\eqno(2.5)$$
Varying with respect to $\s$ and then choosing the gauge $\s=1$, one obtains
the Hamiltonian constraint
$$
\eqalignno{E\id &\ P^2+\l_eZ^2+\l_iV^2+\b U^2&\cr
&+{4\a\over U^2}\left[\l_e\l_iZ^2V^2+\l_eZ^2(Y-W)A+\l_iV^2(X-W)B-
3(X-W)(Y-W)C^2\right]=0,&(2.6)}$$
where
$$P^2=6W^2+3X^2+3Y^2-8WX-8WY+4XY,\qquad C^2=11W^2+4X^2+4Y^2+7XY-13WX-13WY,$$
$$A=4X+3Y-5W,\qquad B=3X+4Y-5W.$$
Variation with respect to $\c$, $\z$ and $\y$ gives rise to the other
\fe
$$\eqalignno{
&2X'+2Y'-3W'+\Big\{{2\a\over U^2}\big[\l_eZ^2(2X+4Y-5W)+\l_iV^2(4X+2Y-5W)+22W^3&\cr
&\quad-2X^3-2Y^3-36W^2X-36W^2Y-12X^2Y-12Y^2X+17WX^2+17WY^2+44XYW\big]\Big\}'&\cr
&\quad={\b\over2}\,U^2+{2\a\over U^2}\big[-\l_e\l_iZ^2V^2+\l_eZ^2(Y-W)A+\l_iV^2(X-W)B-
(X-W)(Y-W)C^2\big],&(2.7)
\cr&&\cr
&X'+2Y'-2W'+\Big\{{4\a\over U^2}\big[\l_eZ^2(2X+2Y-3W)-(X-W)(10W^2+2X^2+5Y^2+6XY&\cr
&\quad-9WX-14WY-\l_iV^2)\big]\Big\}'=\l_eZ^2+\b U^2+{4\a\over U^2}\big[\l_iV^2(X-W)B-(X-W)(Y-W)C^2\big],
&(2.8)\cr&&\cr
&2X'+Y'-2W'+\Big\{{4\a\over U^2}\big[\l_iV^2(2X+2Y-3W)-(Y-W)(10W^2+5X^2+2Y^2+6XY&\cr
&\quad-14WX-9WY-\l_eZ^2)\big]\Big\}'=\l_iV^2+\b U^2+{4\a\over U^2}\big[\l_eZ^2(Y-W)A-(X-W)(Y-W)C^2\big],
&(2.9).}$$
\smallskip

In the variables (2.4), the problem takes the form of a six-dimensional
dynamical system, subject to a constraint. Notice that the function $E$
defined in (2.6) is a constant of the motion of the system (2.5), (2.7)-(2.9), whose
value vanishes by virtue of the Hamiltonian constraint.

\bigskip
\noindent{\it The Einstein limit}
\smallskip
In the Einstein limit $\a=0$ one recovers the results of [3].
We summarize them in terms of the variables introduced above:
when $\a=0$, the dynamical system reduces to eqs. (2.5) and
$$2X'+2Y'-3W'={\b\over2}\,U^2,\qquad X'+2Y'-2W'=\l_eZ^2+\b U^2,\qquad 2X'+Y'-2W'=\l_iV^2+\b U^2,\eqno(2.10)$$
subject to the constraint
$$E=P^2+Z^2+V^2+\b U^2=0.\eqno(2.11)$$
The physical trajectories lie on the four-dimensional hypersurface $E=0$.

The critical points at finite distance correspond to the short radius limit of the
solutions. They lie on the surface $U_0=Z_0=V_0=P_0=0$,
but only points with $X_0=Y_0=W_0$ correspond to regular horizons, while the others give
rise to naked singularities. The
eigenvalues of the linearized equations around the critical points are
$0(3)$, $X_0$, $Y_0$, $W_0$.

The asymptotic properties of the solutions are related to the structure of
the phase space at infinity. This can be investigated defining new variables
$$t={1\over W},\quad x={X\over W},\quad y={Y\over W},\quad u={U\over W},
\quad z={Z\over W},\quad v={V\over W}.\eqno(2.12)$$
In terms of these variables, the field equations at infinity are then
obtained for $t\to0$, and read
$$\eqalignno{&\dot t=-(2v^2+2z^2+{5\over2}\,\b u^2)t,\qquad\qquad\dot u=(1-2v^2-2z^2-{5\over2}\,\b u^2)u,&\cr
&\dot x=z^2+2v^2+2\b u^2-(2v^2+2z^2+{5\over2}\,\b u^2)x\qquad\ \dot z=(x-2v^2-2z^2-{5\over2}\,\b u^2)z,&\cr
&\dot y=2z^2+v^2+2\b u^2-(2v^2+2z^2+{5\over2}\,\b u^2)y,\qquad\dot v=(y-2v^2-2z^2-{5\over2}\,\b u^2)v,&(2.13)}$$
where a dot denotes $t\,d/d\x$.
The critical points at infinity are found at $t_0=0$ and
\medskip
a) $\b u_0^2=\l_iv_0^2=\l_ez_0^2=0$, $x=x_0$, $y=y_0$, with $3x_0^2+3y_0^2+4x_0y_0-8x_0-8y_0+6=0$.

b) $\b u_0^2=\l_iv_0^2=0$, $\l_ez_0^2=1/4$, $x_0=1/2$, $y_0=1$.

c) $\b u_0^2=\l_ez_0^2=0$, $\l_iv_0^2=1/4$, $x_0=1$, $y_0=1/2$.

d) $\b u_0^2=0$, $\l_iv_0^2=\l_ez_0^2=3/16$, $x_0=y_0=3/4$.

e) $\l_ez_0^2=0$, $\l_iv_0^2=-1/3$, $\b u_0^2=2/3$, $x_0=2/3$, $y_0=1$.

f) $\l_iv_0^2=0$, $\l_ez_0^2=-1/3$, $\b u_0^2=2/3$, $x_0=1$, $y_0=2/3$.

g) $\l_iv_0^2=\l_ez_0^2=-1$, $\b u_0^2=2$, $x_0=y_0=1$.

h) $\l_iv_0^2=\l_ez_0^2=0$, $\b u_0^2=2/5$, $x_0=y_0=4/5$.
\medskip
{\noindent Points a) are the endpoints of the hypersurface $U=V=Z=0$, points b)
of the hypersurface $U=V=0$, points c) of the hypersurface $U=Z=0$.}
Clearly, points e)-h) exist only for $\b>0$.

The eigenvalues of the linearized equations around the critical points and their
degeneracies are:
\medskip
a)\quad $0({\it3})$, $1$, $x_0$, $y_0$.

b, c)\quad $-1/2\,({\it3})$, $-1$, $1/2\,({\it2})$.

d)\quad $-3/4\,({\it2})$, $-3/2$, $1/4$, $-{1\over8}(3\pm i\sqrt{15})$.

e), f)\quad $-1\,({\it2})$, $-1/3$, $-2$, $-{1\over6}(3\pm \sqrt{33})$.

g)\quad $-1$, $-2\,({\it3})$, $1\,({\it2})$.

h)\quad $-1\,({\it3})$, $-1/5\,({\it2})$, $-2$.
\medskip
From the study of the eigenvalues and eigenvectors of the linearized equations one can
deduce the structure of phase space at infinity. It results that points a) attract only
unphysical trajectories on the surface at infinity,
while points b)-d) attract only trajectories with $\b=0$. The relevant \cp are therefore
e)-h). Of these, e) attracts trajectories with $\l_e\ge0$, $\l_i<0$, f) attracts trajectories
with $\l_e<0$, $\l_i\ge0$, g) attracts trajectories with both $\l_e<0$, $\l_i<0$ and h)
trajectories with $\l_e\ge0$, $\l_i\ge0$.

The asymptotic behavior of the solutions can be deduced from the location
of the critical points at infinity [2]. Excluding points a) that do not correspond to physical
trajectories, one has, in terms of a radial variable $r$:
\eject
\medskip
\halign{#&\quad#&\quad#\hfil&\qquad#\hfil\cr

&b) &$ds^2\sim -dt^2+dr^2+r^2d\O^2_+$,&$\ef\sim\const.$\cr

&c) &$ds^2\sim -r^2dt^2+r^2dr^2+r^2d\O^2_0$,&$\ef\sim r^2.$\cr

&d) &$ds^2\sim -r\,dt^2+dr^2+r^2d\O^2_+$,&$\ef\sim r.$\cr

&e) &$ds^2\sim -r^2\,dt^2+r^{-2}dr^2+r^2d\O^2_+$,&$\ef\sim\const$.\cr

&f) &$ds^2\sim -r^4\,dt^2+dr^2+r^2d\O^2_-$,&$\ef\sim r^2.$\cr

&g) &$ds^2\sim -r^2\,dt^2+r^{-2}dr^2+d\O^2_-$,&$\ef\sim\const$.\cr

&h) &$ds^2\sim -r^2\,dt^2+r^\mo dr^2+r^2d\O^2_+$,&$\ef\sim r.$\cr}

\medskip

{\noindent We have denoted with $d\O^2_+$ the metric of a unitary 2-sphere, with
$d\O^2_-$ that of a 2-dimensional space of constant negative curvature,
and with $d\O^2_0$ that of a flat 2-plane.}
The solutions ending at points b) are asymptotically flat, those ending at e) asymptotically
anti-de Sitter, while the others have more exotic behavior.

From a \KK point of view, the only solutions with physical relevance are those with $\l_e>0$
and $\ef\to\const$, namely those ending at e), which  have a negative-curvature
internal space. The solutions ending at h) can also have $\l_e>0$, but decompactify for
$r\to\inf$. It follows that, as one could have guessed, the only
significant solutions of this model  are the \as \ads solutions e),
which asymptote to the exact ground state discussed at the end of section 1.

\section{3. The EH-CC-GB phase space}

As discussed in section 1, for a range of values of $\a$ and $\b$  eqs.\ (1.4) admit the ground
state (1.5), and therefore black holes with anti-de Sitter \ab may be expected.
The phase space of the system can be studied by the same methods used in the Einstein
case. However, as usual in the presence of GB terms, some problems arise because of
the poles in the \fe for $U=0$ [7,1].
Special care must therefore be taken when approaching this limit.

Equations (2.7)-(2.9) have to be solved for the variables $X'$, $Y'$ and $W'$
in order to put the system in its canonical form. One can then find the critical points
at finite distance by requiring the vanishing of the derivatives of the fields.
As in the EH-CC case, they lie on the hypersurface $U_0=Z_0=V_0=0$. However,
in the present case, the other variables must satisfy the constraint $W_0=X_0=Y_0$,
or $X_0={4\pm\sqrt5\over5}\,W_0$, $Y_0={4\mp\sqrt5\over5}\,W_0$, in order to avoid
singularities of the field equations. Only the first instance corresponds to regular horizons.
In that case the eigenvalues of the linearized equations near the critical points
are identical to those found in the Einstein limit.

The critical points at infinity are obtained by introducing the variables (2.12) in the
dynamical system and requiring the vanishing of their derivatives as $t\to 0$.
We find the following points:
\bigskip
\def\noah{\noalign{\hrule}}

\halign{\strut#&\vrule\hfil\quad#\hfil\quad&\vrule\hfil\quad#\hfil\quad&
\vrule\hfil\quad#\hfil\quad&\vrule\hfil\quad#\hfil\quad&\vrule\hfil\quad#\hfil\quad&
\vrule\hfil\quad#\hfil\quad\vrule\cr\noalign{\hrule}
&&$x_0$&$y_0$&$\b t_0^2$&$\l_ez_0^2$&$\l_iv_0^2$\cr\noah
&a)&1&1&0&0&0\cr\noah
&b)&1/2&1&0&1/4&0\cr\noah
&c)&1&1/2&0&0&1/4\cr\noah
&e)&2/3&1&${3\pm\sqrt{9-12\g}\over9}$&0&$-{9-12\g\pm2\sqrt{9-12\g}\over9(5-12\g)}$\cr\noah
&f)&1&2/3&${3\pm\sqrt{9-12\g}\over9}$&$-{9-12\g\pm2\sqrt{9-12\g}\over9(5-12\g)}$&0\cr\noah
&g)&$1$&1&$1\pm\sqrt{1-4\g}$&$-1$&$-1$\cr\noah
&h)&4/5&4/5&${5\pm\sqrt{25-60\g}\over25}$&0&0\cr\noah
&i)&2/3&1&0&1/3&0\cr\noah
&l)&1&2/3&0&0&1/3\cr\noah
&m$_\pm$)&${4\pm\sqrt5\over5}$&${4\mp\sqrt5\over5}$&0&0&0\cr\noah
 }
\bigskip
{\noindent where $\g=\a\b$. Of course, the critical points e), f) can only exist if $\g\le3/4$,
the points g) if $\g\le1/4$ and the points h) if $\g\le5/12$. For these values of $\g$, the
location of the critical points is rather similar to that obtained in the $\a=0$ limit,
except for the point d), that has disappeared and the new points i)-m$_\pm$), that are typical of
the GB theory [1]. In the limit $\g\to0$ one recovers  the  \cp of the EH-CC model.

A stronger similarity is however present with the phase space of the EH-GB limit $\b=0$.
In fact, one has exactly the same critical points as for nonvanishing $\b$, with only the
values of  $u_0$, $z_0$ and $v_0$ shifted. As we shall see, however, the nature of the
critical points may be different in the two cases. Moreover, the limit $\b\to0$ is not trivial.

In the following table are the eigenvalues of the linearized equations near the \cp
\bigskip
\halign{\strut#&\vrule\hfil\quad#\hfil\quad&
&\vrule\hfil\quad#\hfil\quad\vrule\cr\noalign{\hrule}
&&Eigenvalues (with degeneracy)\cr\noah
&a)&$0\,({\it2})$, $1\,({\it3})$, $4\g$\cr\noah
&b), c)&$-{1\over2}$, $-{1\over2}+4\g$, ${1\over2}\,({\it2})$,
$-{3-8\g\pm\sqrt{1+16\g-32\g^2}\over4}$\cr\noah
&e), f)&$-1\,({\it2})$, $-2$, $-{1\over3}$, $-{3\pm\sqrt{33-32\g}\over6}$\cr\noah
&g)&$-1$, $-2\,({\it2})$, 1, $-\ha(1\pm\sqrt\S)$\cr\noah
&h)&$-1\,({\it3})$, $-2$, $-{1\over5}\,({\it2})$\cr\noah
&i), l)&$-{2\over3}$, $-{1\over3}$, $-1$, ${1\over3}\,({\it3})$\cr\noah
&m$_\pm$)&$-{2\over 3}\,({\it2})$, 0, ${1\over3}$, ${2\pm3\sqrt{5}\over15}$\cr\noah
}
\bigskip
{\noindent In the last eigenvalues of g), $\S$ is a cumbersome function of $\g$. It turns out
that the real part of both eigenvalues is negative for $\g<-3/4$, while for $-3/4<\g<1/4$,
one of the eigenvalues has positive real part.}

From the study of the linearized equations, one can deduce that point a)
attracts only trajectories lying on the surface at infinity, while
the points i), l), m$_\pm$) attract only points on the hypersurface
$U=0$, which corresponds to the limit $\a\to\inf$ of pure GB gravity [1].
Therefore, these points are not of interest for our problem.
Moreover, points b), c) are endpoints only of trajectories with $\b=0$.
The other points can attract trajectories with nonvanishing $\b$. In particular,
e) attracts trajectories with $\l_i=0$, f)  trajectories with $\l_e=0$, g) trajectories
with $\l_e<0$, $\l_i <0$, and h) trajectories with any values of $\l_e$, $\l_i$.

The \cp a)-h) generalize those found in the EH-CC case, and have
the same \ab as the corresponding points. We do not discuss the \ab of the new points
i), l), m$_\pm$), since they correspond to the limit $\a\to\inf$.

Of particular interest are the solutions that end at the critical point e). These
asymptote to the exact ground state solution discussed in section 1, namely $AdS^4\times S^2$,
if $\a>0$ and $5/12<\g<3/4$ or $\a<0$ and $\b>0$, or to $AdS^4\times H^2$, if $\a>0$ and
$\g<5/12$.
Contrary to the EH-CC case, solutions with \ads
asymptotics can therefore exist also if $\b<0$.
In the present \coo they take the form
$$ds^2=-\left({r^2\over|\L_e|}+1\right)dt^2+\left({r^2\over|\L_e|}+1\right)^\mo dr^2+r^2d\O^2_+,
\qquad\ef=|\L_i|,\eqno(3.1) $$
where $\L_e$, $\L_i$ are given by (1.5).

Also interesting is the solution g), that asymptotes to the exact solution $AdS^2\times H^2
\times H^2$.
Its four-dimensional section is analogous to a Bertotti-Robinson metric. The other
solutions have less common asymptotical behavior.

The structure of the solutions of the EH-GB-CC model is more complicated than that of pure
EH-CC, although less \cp are available for $\b\ne0$. The trajectories start at the points
$U=V=Z=0$, $W=X=Y$ and can terminate at one of the points e)-h), depending on the
value of $\l_e$, $\l_i$ and on the value of $\g$.
From the \KK point of view, the relevant solutions are those with $\l_e>0$ and
$\ef\to$ const at infinity. As in the $\b=0$ case, these are solutions e), but if $\b\ne0$,
the internal space can be flat (if $\g=3/4$), or compact, if $5/12<\g<3/4$ or $\a<0$, $\b>0$.

The critical point e) attracts a three-dimensional bunch of trajectories, of which a
two-dimensional subset has a regular horizon. It is expected therefore that these solutions
constitute a two-parameter family. An explicit example for special values of $\a$ and $\b$ is
given in ref.\ [8].

\section{4. Conclusions}

We have shown that regular black holes in EH-CC-GB theory have the same \ab as the maximally symmetric
ground states. In contradistinction to the EH-CC case, the internal space can have  positive,
negative or vanishing curvature.

Although the calculation are easier in our six-dimensional model,
we believe that the situation is essentially unchanged in higher dimensions, except that in case
of three or more internal dimensions ground states with flat spacetime can exist for some values
of the parameters.

\bigskip

\section{Appendix A}
In the Einstein limit, $\a=0$, the \fe reduce to
$$\c''=2\l_e e^{2\z}+2\l_ie^{2\y}+{5\b\over2}\,e^{2\c},\qquad
\z''=\l_e e^{2\z}+2\l_ie^{2\y}+2\b\,e^{2\c},\qquad
\y''=2\l_e e^{2\z}+\l_ie^{2\y}+2\b\,e^{2\c},\eqno(A.1)$$
subject to the constraint
$$6\c'^2+3\z'^2+3\y'^2-8\c'\z'-8\c'\y'+4\z'\y'+
\l_ee^{2\z}+\l_ie^{2\y}+\b e^{2\c}=0.\eqno(A.2)$$

Some exact solutions of the system (A.1)-(A.2) can be found in special cases. The
limit $\b=0$ has been discussed in [1,2]:
when $\l_i=0$ one obtains  the \sch metric with constant scalar field,
$$ds^2=-\left(1-{2m\over r}\right)dt^2+\left(1-{2m\over r}\right)^\mo dr^2
+r^2d\O^2_+,\qquad\ef=\const,$$
with $m$ a free parameter. This is a special case of solutions b) of section 2.

For $\l_e=0$, one has instead a solution of the form
$$ds^2=-r^2\left(1-{2m\over r}\right)dt^2+r^2\left(1-{2m\over r}\right)^\mo dr^2+
r^2d\O^2_0,\qquad\ef=r^2,$$
which corresponds to the \ab c).

Finally, if $\y'=\z'$, one has
$$ds^2=-r\left({4\over27}-{2m\over r^{3/2}}\right)dt^2+\left({4\over27}-{2m\over r^{3/2}}
\right)^\mo dr^2+r^2d\O^2,\quad\ef=r,$$
which corresponds to d).

In the following, it will be useful to write the solutions in their six-dimensional
Schwarzschild-like form
$$ds^2=-e^{2\l}dT^2+e^{-2\l}dR^2+e^{2\r}d\O^2_e+e^{2\s}d\O^2_i,\eqno(A.3)$$
where
$$e^{2\l}=e^{4\z+4\y-6\c},\qquad\e^{2\r}=e^{2\c-2\z},\qquad\e^{2\s}=e^{2\c-2\y}.\eqno(A.4)$$

In these \coo the previous solutions read respectively
$$ds^2=-\left(1-{2M\over R}\right)dT^2+\left(1-{2M\over R}\right)^\mo
dR^2+R^2d\O^2_++c^2d\O^2_0,$$
$$ds^2=-\left(1-{2M\over R}\right)dT^2+\left(1-{2M\over R}\right)^\mo
dR^2+c^2d\O^2_0+R^2d\O^2_,$$
$$ds^2=-\left(1-{2M\over R^3}\right)dT^2+\left(1-{2M\over R^3}\right)^\mo dR^2
+{R^2\over3}\,(d\O_i^2+d\O_e^2),$$
with $c=\const$.

We pass now to consider the case when $\b\ne0$.
\bigskip
{\noindent 1) $\l_e=0$, $\l_i<0$, $\y'=\c'$.}
\smallskip
The field equations (A.1) reduce to
$$e^{2\y}={\b\over2}e^{2\c},\qquad\c''={3\over2}\,\b\, e^{2\c},\qquad\z''={2\over3}\c''.$$
Integrating,
$$e^{2\c}={2\over3\b}\ {a^2\over\sinh^2a\x},\qquad e^{2\z}=A e^{2(2\c+b\x)/3},$$
for constant $a$ and $b$.
Substituting in the constraint (A.2) and requiring the presence of a regular horizon,
one obtains the condition $b=a$.
Then, defining $R=[c/(1-e^{2a\x})]^{1/3}$, with $c$ a positive constant, one gets
$$e^{2\c}=AR^3(R^3-c),\quad e^{2\z}=BR(R^3-c),$$
where $A=8a^2/3\b c^2$ and $B$ is an integration constant. Finally, choosing $B=A$,
rescaling the time coordinate, and defining $M=\b c/12$, one obtains
$$e^{2\l}={\b\over6}\,R^2-{2M\over R},\qquad e^{2\r}=R^2,\qquad e^{2\s}={2\over\b}.\eqno(A.5)$$
In four-dimensional coordinates the solution reads
$$ds^2=-\left({\b^2\over12}\,r^2-{2m\over r}\right)dt^2+\left({\b^2\over12}\,r^2-{2m\over r}\right)^\mo dr^2
+r^2d\O_0^2,\qquad\ef={2\over\b},$$
and has therefore the \ab e).

The solution (A.5) can also be generalized to the case of positive $\l_e$,
although the solution is not trivial in the coordinates $\z$, $\y$, $\c$. It reads
$$-\left({\b\over6}\,R^2+1-{2M\over R}\right)dT^2+\left({\b\over6}\,R^2+1-{2M\over R}\right)^\mo
dR^2+R^2d\O_+^2+{2\over\b}\,d\O_-^2,$$
and is the direct product of $AdS^4$ with a two-dimensional space of constant negative curvature
$H^2$.
\bigskip
{\noindent 2) $\l_i=0$, $\l_e<0$, $\z'=\c'$}.
\smallskip
This system is identical to that of the previous case, after interchanging $\z$ and $\y$
(and hence $\r$ and $\s$).
Proceeding as before, one obtains
$$e^{2\l}={\b\over6}\,R^2-{2M\over R},\qquad e^{2\r}={2\over\b},\qquad e^{2\s}=R^2.\eqno(A.6)$$
In four-dimensional coordinates (2.1), the solution reads
$$ds^2=-r^4\left({1\over3}-{2m\over r^3}\right)dt^2+\left({1\over3}-{2m\over r^3}\right)^\mo dr^2+r^2d\O_-^2,\qquad\ef=r^2,$$
and has therefore the \ab f).

In analogy with the previous case, the solution (A.6) can be generalized to positive $\l_i$,
$$-\left({\b\over6}\,R^2+1-{2M\over R}\right)dT^2+\left({\b\over6}\,R^2+1-{2M\over R}\right)^\mo
dR^2+{2\over\b}\,d\O_-^2+R^2d\O_+^2.$$

\bigskip
{\noindent 3) $\l_e=\l_i=0$, $\y'=\z'$.}
\smallskip
In this case, the field equations reduce to
$$\c''={5\over2}\,\b\, e^{2\c},\qquad\z''={4\over5}\c''.$$
Integrating,
$$e^{2\c}={2\over5\b}\ {a^2\over\sinh^2a\x},\qquad e^{2\z}=e^{2\y}=A e^{2(4\c+b\x)/5},$$
for constant $a$ and $b$.
Substituting in the constraint, and requiring the presence of a regular horizon,
one obtains the condition $b=a$.
Defining $R=[c/(1-e^{2a\x})]^{1/5}$, for constant $c$, one gets
$$e^{2\c}=AR^5(R^5-c),\quad e^{2\z}=BR^3(R^5-c),$$
where $A=8a^2/5\b c^2$ and $B$ is an integration constant. Finally, choosing $B=A$, rescaling
$T$, and defining $M=5\b c/4$, one obtains
$$e^{2\l}={5\b\over2}\,R^2-{2M\over R^3},\qquad e^{2\r}=e^{2\s}=R^2.\eqno(A.7)$$

In four-dimensional coordinates the solution can be written
$$-r^2\left(10\b -{2m\over r^{5/2}}\right)dt^2+{1\over r}\left(10\b -{2m\over r^{5/2}}\right)^\mo dr^2+r^2d\O_0^2,\qquad\ef=r,$$
and has therefore the \ab h). In this form, the solution was obtained in [2].

The solution (A.7) can also be generalized to the case of positive $\l_e$ and $\l_i$,

$$-\left[{5\b\over2}\,R^2+1-{2M\over R^3}\right]dT^2+\left[{5\b\over2}\,R^2+1-{2M\over R^3}
\right]^\mo dR^2+{R^2\over3}(d\O_+^2+d\O_+^2).$$
In this form it generalizes the six-dimensional Tangherlini-\ads metric [9], with $S^4$
replaced by $S^2\times S^2$.

\bigskip
{\noindent 4) $\l_e<0$, $\l_i<0$, $\y'=\z'=\c'$}.
\medskip
The \fe yield
$$\c''={\b\over2}\,e^{2\c},\qquad e^{2\z}=e^{2\y}={\b\over2}\,e^{2\c},$$
and hence
$$e^{2\c}={2\over\b}\ {a^2\over\sinh^2a\x}.$$
Defining a variable $R=2a/(1-e^{2a\x})$, one gets
$$e^{2\l}={\b\over2}R(R-2M),\qquad e^{2\r}=e^{2\s}={2\over\b},\eqno(A.8)$$
where $M=a$. The metric is clearly the direct product $AdS^2\times H^2\times H^2$.
After dimensional reduction,
$$ds^2=-r(r-2m)dt^2+{dr^2\over r(r-2m)}+{4\over\b^2}\,d\O_-^2,\qquad\ef={2\over\b},$$
which corresponds to the \ab g).

\bigbreak
\section{Appendix B}
Some of the exact solutions of the previous appendix can be extended to the GB case.
We write them in six-dimensional form, since the duality is more apparent.

\medskip
\halign{#&\quad#&\quad#\hfil&\qquad#\hfil\cr

&e) &$ds^2=-\left(|\L_e|\,r^2+1\right)dt^2+\left(|\L_e|\,r^2+1\right)^\mo dr^2+r^2d\O^2_++{1\over|\L_i|}\,d\O^2_-$,\qquad or\cr
&&$ds^2=-|\L_e|\,r^2\,dt^2+\left(|\L_e|\,r^2\right)^\mo dr^2+r^2d\O^2_0+{1\over|\L_i|}\,d\O^2_-$.\cr

&f) &$ds^2=-\left(|\L_e|\,r^2+1\right)dt^2+\left(|\L_e|\,r^2+1\right)^\mo dr^2+{1\over|\L_i|}\,d\O^2_-+r^2d\O^2_+$,\qquad or\cr
&&$ds^2=-|\L_e|\,r^2\,dt^2+\left(|\L_e|\,r^2\right)^\mo dr^2+{1\over|\L_i|}\,d\O^2_-+r^2d\O^2_0$.\cr

&g) &$ds^2=-\left({r^2\over\D}-m\right)dt^2+\left({r^2\over\D}-m\right)^\mo dr^2+\D\,d\O^2_-+\D\,d\O^2_-\,.$\cr

&h) &$ds^2=-{r^2\over\G}\,dt^2+{\G\over r^2}\,dr^2+r^2d\O^2_0+r^2d\O^2_0.$\cr}
\noindent{where $\L_e$ and $\L_i$ are given by (1.5) and}
$$\D={1\pm\sqrt{1-4\a\b}\over2\b},\qquad\G={5(1\pm\sqrt{1-12\a\b/5})\over2\b}.$$

An interesting two-parameter exact solution for special values of $\a$ and $\b$
has recently been given in [8].
\bigskip
\beginref
\ref [1] S. Mignemi, \grq{0607005}.
\ref [2] S. Mignemi and D.L. Wiltshire, \CQG{6}, 987 (1989).
\ref [3] D.L. Wiltshire, \PR{D44}, 1100 (1991).
\ref [4] D. Lovelock, \JMP{12}, 498 (1971);
B. Zwiebach, \PL{B156}, 315 (1985);
B. Zumino, \PRep{137}, 109 (1986).
\ref [5] J. Madore, \PL{A110}, 289 (1985); \PL{A111}, 283 (1985);
F. M\"uller-Hoissen, \PL{B163}, 106 (1985); \CQG{3}, L133 (1986);
S. Mignemi, \MPL{A1}, 337 (1986).
\ref [6] D.G. Boulware and S. Deser, \PRL{55}, 2656 (1985);
J.T. Wheeler, \NP{B268}, 737 (1986);
D.L. Wiltshire, \PL{B169}, 36 (1986);
R.-G. Cai, \PR{D65}, 084014 (2002);  R.-G. Cai and Q. Guo,
\PR{D69}, 104025 (2004).
\ref [7] M. Melis and S. Mignemi, \CQG{22}, 3169 (2005); \PR{D73}, 083010 (2006).
\ref [8] H. Maeda and N. Dadhich, \PR{D74}, 021501 (2006).
\ref [9] D. Xu, \CQG{5}, 871 (1988).

\endref
\end